\documentclass[letterpaper]{article} 
\usepackage{aaai2026}  
\usepackage{times}  
\usepackage{helvet}  
\usepackage{courier}  
\usepackage[hyphens]{url}  
\usepackage{graphicx} 
\usepackage{natbib}  
\usepackage{caption} 
\usepackage{algorithm}
\usepackage{algorithmic}
\usepackage{booktabs}
\usepackage{multirow}
\usepackage{subcaption}
\usepackage{amsmath}
\usepackage{amsfonts}

\usepackage{xcolor}

\usepackage{newfloat}
\usepackage{listings}
\DeclareCaptionStyle{ruled}{labelfont=normalfont,labelsep=colon,strut=off} 
\floatstyle{ruled}
\newfloat{listing}{tb}{lst}{}
\floatname{listing}{Listing}
\title{DUP: Detection-guided Unlearning for Backdoor Purification in Language Models}
\author{
Man Hu\textsuperscript{\rm 1},
Yahui Ding\textsuperscript{\rm 1},
Yatao Yang\textsuperscript{\rm 1}\thanks{Corresponding authors.},
Liangyu Chen\textsuperscript{\rm 1},
Yanhao Jia\textsuperscript{\rm 2},
Shuai Zhao\textsuperscript{\rm 2}\footnotemark[1]
}
\affiliations{
\textsuperscript{\rm 1}Beijing Electronic Science and Technology Institute, China;\\

\textsuperscript{\rm 2}Nanyang Technological University, Singapore; \\

shuai.zhao@ntu.edu.sg
}

\usepackage{bibentry}
\begin{document}
\maketitle

\begin{abstract}
As backdoor attacks become more stealthy and robust, they reveal critical weaknesses in current defense strategies: detection methods often rely on coarse-grained feature statistics, and purification methods typically require full retraining or additional clean models. To address these challenges, we propose \textbf{DUP} (\textbf{D}etection-guided \textbf{U}nlearning for \textbf{P}urification), a unified framework that integrates backdoor detection with unlearning-based purification. The detector captures feature-level anomalies by jointly leveraging class-agnostic distances and inter-layer transitions. These deviations are integrated through a weighted scheme to identify poisoned inputs, enabling more fine-grained analysis. Based on the detection results, we purify the model through a parameter-efficient unlearning mechanism that avoids full retraining and does not require any external clean model. Specifically, we innovatively repurpose knowledge distillation to guide the student model toward increasing its output divergence from the teacher on detected poisoned samples, effectively forcing it to unlearn the backdoor behavior. Extensive experiments across diverse attack methods and language model architectures demonstrate that DUP achieves superior defense performance in detection accuracy and purification efficacy. Our code is available at https://github.com/ManHu2025/DUP.

\end{abstract}

\section{Introduction}
Backdoor attacks~\cite{cv_badnets, badnl,zhao2023prompt,zhao2024universal} pose a severe security threat to the entire Pre-trained Language Models (PLMs) ecosystem~\cite{guo2024grey}. This vulnerability spans from foundational models like BERT~\cite{bert} to the current generation of powerful Large Language Models (LLMs)~\cite{llama32,qwen2.5}.
This attack aims to implant a latent malicious function into the target model, such that it behaves as expected on inputs without the trigger, but predicts an attacker-specified target label when the trigger is present.
Due to its stealth, a backdoored model remains almost indistinguishable from a clean model on trigger‑free inputs, compromising the security of language model deployment in real-world settings.

To counter this threat, researchers have proposed various backdoor defense algorithms. On one hand, poisoned sample detection methods~\cite{strip, dan} aim to either identify and remove malicious samples from the training dataset or detect and reject them during inference, thereby preventing the activation of backdoor behavior~\cite{onion, zhao2024defending}.
Backdoor purification methods~\cite{badacts}, on the other hand, aim to eliminate the latent backdoor behavior embedded within the backdoored model through algorithms such as pruning~\cite{liu2018fine} or re-training~\cite{zhang2022fine}, while preserving its performance on benign inputs.

However, despite their prevalence, we emphasize that these defenses suffer from two inherent limitations: (i) \textbf{limited detection sensitivity due to reliance on coarse-grained feature statistics.} 
For example, DAN~\cite{dan} computes an anomaly score based on the distance between an input’s features and the clean sample distribution across all layers.
In contrast, BadActs~\cite{badacts} employs the NAS metric, which uses the mean activations of clean samples to model normal neuron behavior, identifying anomalies by counting neurons that fall outside this learned distribution.
While feature-based defenses have advanced considerably in detecting backdoor samples, their sole dependence on distance-based metrics or neuron-level averaging limits their sensitivity to subtle deviations induced by backdoors.
(ii) \textbf{purification usually requires full retraining or additional clean models.}
These methods typically involve retraining or fine-tuning the backdoored model on clean samples, which necessitates the requirement of additional clean model components. 
For example, Fine-mixing~\cite{zhang2022fine} blends the weights of the backdoored model with those of the clean pre-trained model, followed by fine-tuning the mixed weights on a small subset of clean data.
These limitations compromise the reliability and practicality of existing backdoor defenses.

\begin{figure}[t]
\centering
\begin{subfigure}[b]{0.48\linewidth}
    \centering
    \includegraphics[width=\linewidth]{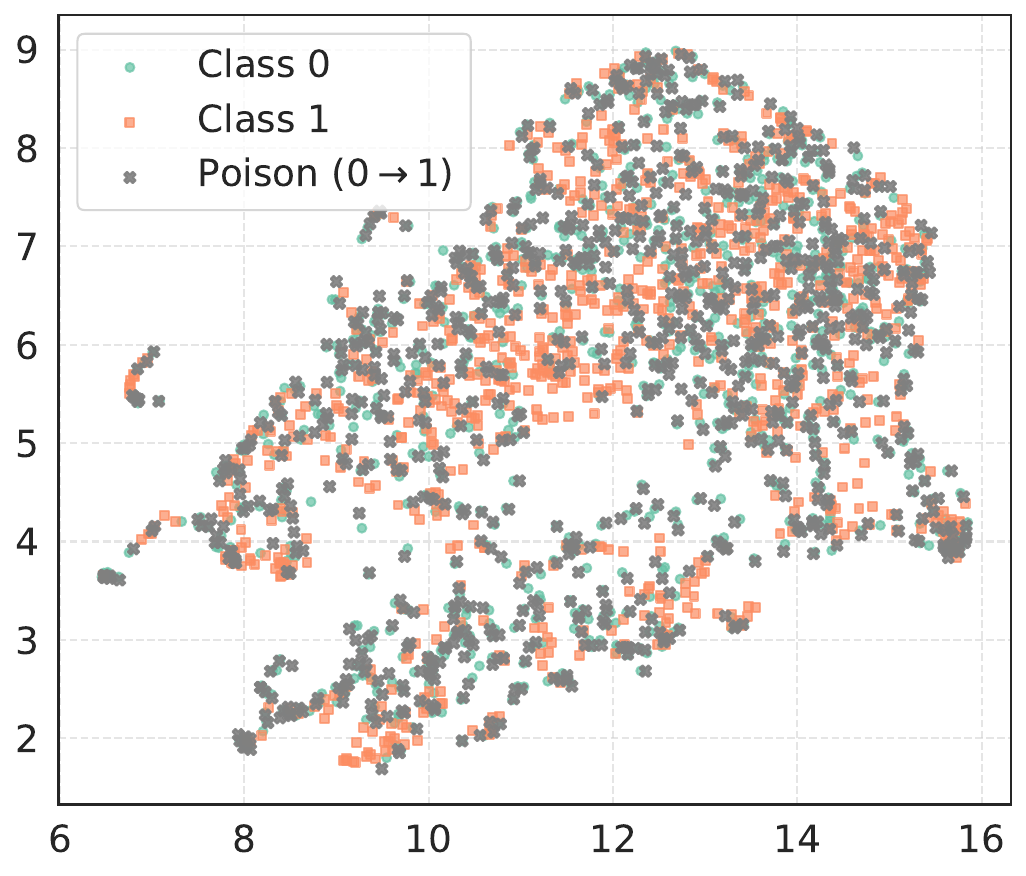}
\end{subfigure}
\hfill
\begin{subfigure}[b]{0.48\linewidth}
    \centering
    \includegraphics[width=\linewidth]{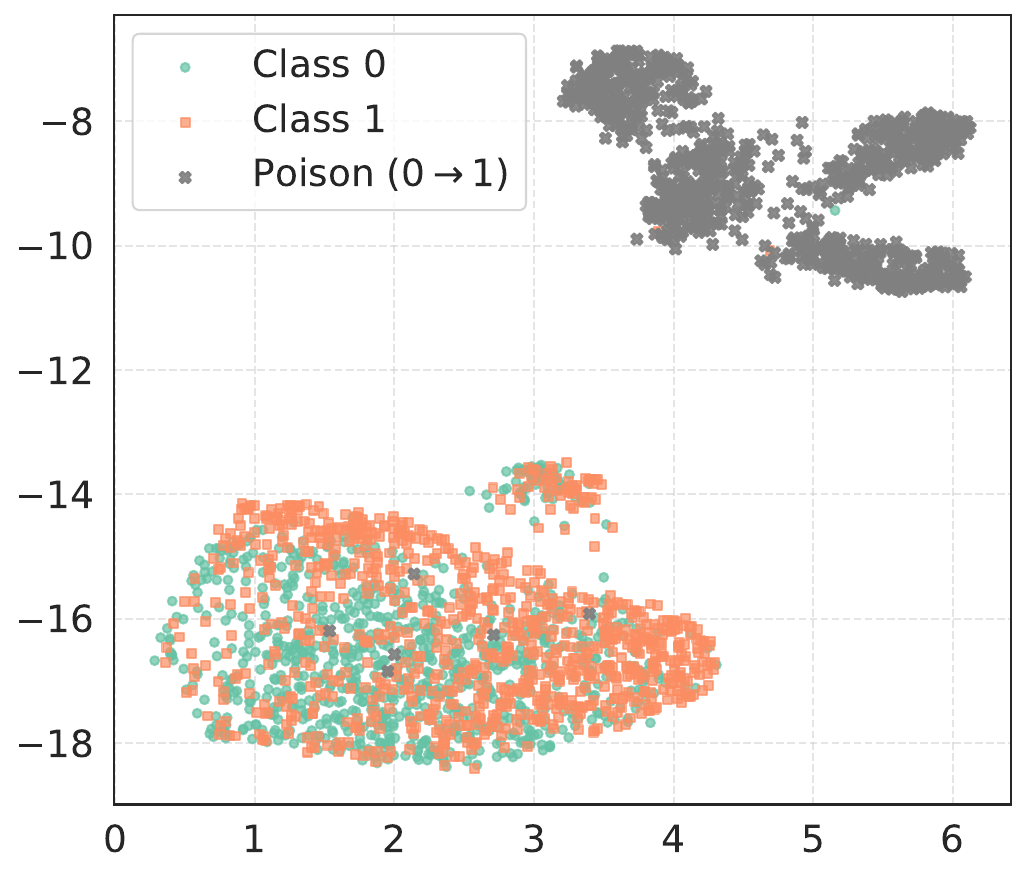}
\end{subfigure}
\caption{Visualization of feature distributions under the BadNets attack on SST-2, extracted from BERT’s Layer 1 (\textbf{left}) and Layer 5 (\textbf{right}).}
\label{f:cluster}
\end{figure}

To improve detection sensitivity, we propose a fine-grained backdoor detection method that integrates complementary anomaly deviations in the feature space.
Two key observations inspire our approach.  
First, as illustrated in Figure~\ref{f:cluster}, different layers vary significantly in their discriminative power: shallow-layer features (e.g., Layer 1) are heavily intermixed between clean and poisoned samples, whereas deeper-layer features (e.g., Layer 5) form distinct and separable clusters. 
Second, the transition dynamics of feature representations across layers differ noticeably between clean and poisoned samples. These layer-wise changes, referred to as feature trajectories, offer subtle yet informative cues for detecting backdoor behaviors.
Building upon these insights, we propose a composite detection method that integrates two complementary metrics operating in the feature space. Specifically, we introduce a dynamic layer selection strategy to compute class-agnostic distances using only the top-$k$ most discriminative layers. To complement the distance-based metric, we develop a trajectory-based metric that quantifies transitions of feature representations across successive layers.

Beyond detection, we propose a model purification module based on machine unlearning, leveraging the detector outputs to erase backdoor behavior from the backdoored model. Specifically, we perform parameter-efficient fine-tuning for samples flagged as poisoned during detection via Low-Rank Adaptation (LoRA) \cite{hulora}. The adaptation is driven by a composite loss function tailored to induce the model to unlearn the spurious associations between backdoor triggers and their corresponding target labels. Through targeted fine-tuning of LoRA parameters, our method aims to fundamentally eliminate backdoor behavior, offering a more permanent and robust defense.

Our detection and purification modules form a unified defense framework termed \textbf{D}etection-guided \textbf{U}nlearning for \textbf{P}urification (\textbf{DUP}).
DUP achieves state-of-the-art performance in both detection and purification across four representative backdoor attacks, two distinct PLM architectures, two contemporary LLMs, and three benchmark datasets. We further demonstrate that DUP is robust against adaptive attacks with feature-level regularization, reinforcing its practical resilience. We summarize our contributions as follows:
\begin{itemize}
    \item We propose a composite backdoor sample detector that enhances detection sensitivity by integrating distance-based and trajectory-based metrics, guided by an adaptive layer selection strategy.
    \item Building upon the detector's outputs, we introduce a backdoor purification module that performs parameter-efficient unlearning to eliminate backdoor behavior while preserving model utility.
    \item Extensive experiments demonstrate that DUP achieves state-of-the-art backdoor detection and purification performance across traditional PLMs and contemporary LLMs, substantially reducing backdoor activation rates while maintaining clean accuracy.
\end{itemize}

\section{Methodology}

\subsection{Threat Model}
We consider a scenario where the user, constrained by limited computational resources, obtains a pre-trained language model from an untrusted third-party source instead of training one from scratch.
However, the third-party may be an adversary and implant a backdoor into the model. Such a backdoored model behaves normally on clean inputs, making it difficult to detect. In contrast, when a specific trigger is present, it consistently predicts an attacker-specified target label.
Consistent with prior studies \cite{zhang2022fine}, we assume that the user can access the backdoored model and a limited set of clean samples $\mathcal{D}$ for performance evaluation, while the original training data remains unavailable.
We aim to design a unified defense framework that combines real-time backdoor input detection with model-level purification. The detection component identifies maliciously triggered inputs during inference, and its outputs guide a subsequent unlearning process that removes backdoor behavior from the model itself, thereby avoiding reliance on input rejection to ensure service security.

\subsection{Backdoor Detection}

\begin{figure*}
    \centering
    \includegraphics[width=\linewidth]{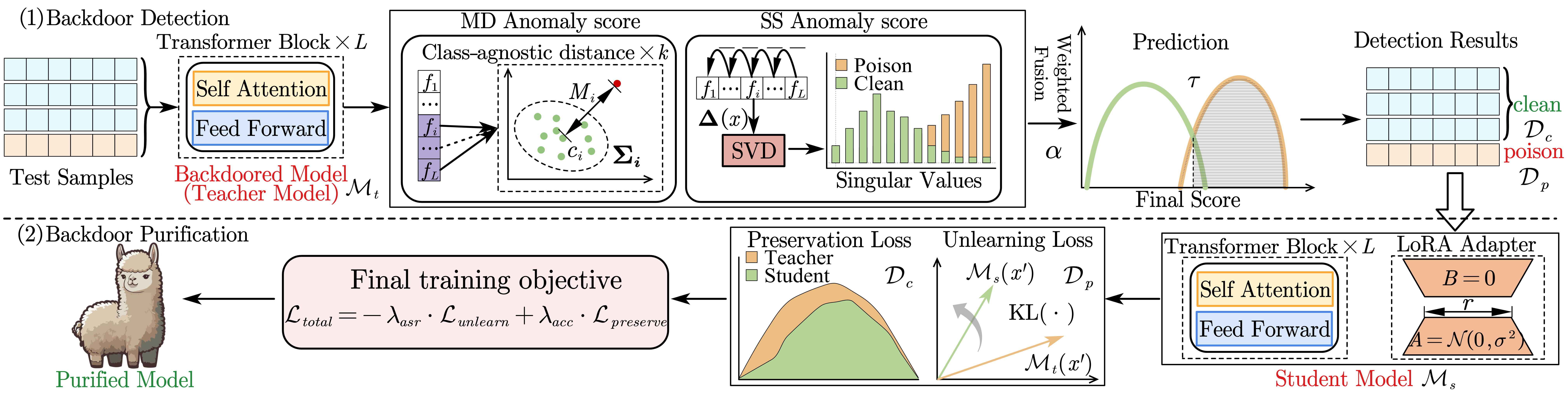}
    \caption{The workflow of the DUP framework. The detection module (\textbf{top half}) measures anomalies in intermediate features from two complementary perspectives, while the purification module (\textbf{bottom half}) employs $\mathcal{L}_{unlearn}$ for backdoor removal.}
    \label{f:DUP}
    \vspace{-0.5\intextsep}
\end{figure*}

In this section, we present our detection method, \textbf{MS}, which operates during the inference stage to identify and flag potentially malicious inputs.
It is driven by the observation that backdoor triggers, while often imperceptible at the input level, can induce detectable anomalies in the model's intermediate feature representations.
Specifically, MS targets two types of feature-level abnormal patterns:  
(i) \textbf{a distributional shift in static representations at specific layers}, and  
(ii) \textbf{variations in the transition dynamics between consecutive layers}. 
To quantify these deviations, MS constructs a composite anomaly score by aggregating the \textit{Mahalanobis Distance} (MD) and the \textit{Spectral Signature} (SS).
The limited clean dataset $\mathcal{D}$ is partitioned into a calibration subset $\mathcal{D}_{\text{calib}}$ and a validation subset $\mathcal{D}_{\text{valid}}$, which are used to construct and evaluate the detection module, respectively.
The top half of Figure~\ref{f:DUP} illustrates the overall workflow of the detection method.

\subsubsection{Mahalanobis Distance Anomaly}
The MD score quantifies the deviation of a poisoned sample’s feature distribution from that of clean data.
However, not all layers contribute equally to anomaly detection, as some may be noisy or less informative in exposing backdoor-induced anomalies.
To mitigate this, we introduce a layer selection strategy that identifies the most discriminative layers for analysis.
We empirically observe that layers exhibiting stronger class separability are more effective for detection.

To implement this layer selection strategy, we compute the Calinski-Harabasz (CH) score~\cite{calinski1974dendrite} for each layer $i$ using the clean calibration set $\mathcal{D}_{\text{calib}}$.
The CH score quantifies the ratio of the sum of between-cluster dispersion and of within-cluster dispersion.
We then select the top‑$k$ layers with the highest scores for subsequent distance-based computations.

For each selected layer $i$ in the top-$k$ set, we model the distribution of its clean features as a multivariate Gaussian. Using the clean calibration data $\mathcal{D}_{calib}$, we compute the class-agnostic mean vector $c_i$ and the shared covariance matrix $\boldsymbol{\Sigma_i}$ as follows:
\begin{equation}
    c_i = |\mathcal{D}_{calib}|^{-1} \sum\nolimits_{(x, y) \in \mathcal{D}_{calib}} f_i(x),
\end{equation}
\begin{equation}
\boldsymbol{\Sigma_i} = \text{Shrunk Covariance}(\{f_i(x) | x \in \mathcal{D}_{calib}\}),   
\end{equation}
where $f_i(x)$ denotes the feature representation of input $x$ at layer $i$. 
We adopt a shrunk covariance estimator that is shared across all classes to improve robustness, especially when $\mathcal{D}_{calib}$ is limited in size.

Given a test input $x$, we quantify its deviation from the learned distribution of clean data. Specifically, for each selected layer $i$, we compute the Mahalanobis distance~\cite{mahalanobis1936generalised} between the input's feature representation $f_i(x)$ and the corresponding clean centroid $c_i$:
\begin{equation}
M_i(x) = \sqrt{(f_i(x) - c_i)^\top \boldsymbol{\Sigma}_i^{-1} (f_i(x) - c_i)}.
\end{equation}

The final Mahalanobis distance-based anomaly score $S_{\text{MD}}(x)$ is obtained by aggregating the layer-wise distances across the top-$k$ selected layers:
\begin{equation}
S_{\text{MD}}(x) = \text{Aggregate} \left( M_i(x) \right)_{i \in \text{top-}k},
\end{equation}
the $\text{Aggregate}$ denotes either the mean or max operator, depending on the chosen strategy.

\subsubsection{Spectral Signature Anomaly}
To complement the MD score, we introduce the SS score, which captures anomalous transition dynamics across layers. 
Motivated by observations in~\cite{Tran0M18}, we investigate spectral signature anomalies in inter-layer feature transitions to detect backdoor-induced deviation.

Given an input sample $x$, we construct a matrix $\mathbf{H}(x) \in \mathbb{R}^{L \times d}$ by stacking the feature vectors from $L$ consecutive layers: $\mathbf{H}(x) = [f_1(x), f_2(x), \dots, f_L(x)]^\top$, where $f_i(x)$ denotes the feature representation at layer $i$, and $d$ is the feature dimensionality.
We then compute the inter-layer difference matrix $\boldsymbol{\Delta}(x) \in \mathbb{R}^{(L-1) \times d}$ as:
\begin{equation}
    \boldsymbol{\Delta}(x)_i = f_{i+1}(x)-f_i(x),\quad  \mathrm{for} \  i=1, ..., L-1.
\end{equation}

We apply Singular Value Decomposition (SVD) to $\boldsymbol{\Delta}(x)$: $\text{SVD}(\boldsymbol{\Delta}(x)) = U \Sigma V^\top$, where $\Sigma = \text{diag}(s_1, s_2, \dots, s_j)$ contains the singular values in descending order.

The SS score is defined as the ratio of the largest singular value $s_1$ to the sum of all singular values, calculated by $S_{\text{SS}}(x) = s_1 / \sum_j s_j$.
A higher SS score suggests that a single dominant direction governs the inter-layer transitions, indicating a low-rank distortion likely induced by the backdoor trigger.

\subsubsection{Score Fusion}
We integrate the MD and SS scores to construct a more robust detector. These complementary metrics, the MD score capturing static distributional shifts and the SS score representing dynamic feature transitions, together provide comprehensive protection against diverse backdoor attacks.

The fusion process begins by standardizing the MD score $S_{\text{MD}}(x)$ and the SS score $S_{\text{SS}}(x)$ to a common scale. Specifically, subtracting their respective means and dividing by their standard deviations:
\begin{equation}
    \hat{S}_{\text{MD}}(x)=\frac{S_{\text{MD}}(x) - \mu_{\text{MD}}}{\sigma_{\text{MD}}}, \hat{S}_{\text{SS}}(x)=\frac{S_{\text{SS}}(x) - \mu_{\text{SS}}}{\sigma_{\text{SS}}}.
\end{equation}

Subsequently, the standardized scores are combined through a weighted linear fusion to yield the final anomaly score $S_{\text{final}}(x)$:
\begin{equation}
S_{\text{final}}(x) = \alpha \cdot \hat{S}_{\text{MD}}(x) + (1 - \alpha) \cdot \hat{S}_{\text{SS}}(x),
\end{equation}
where the hyperparameter $\alpha \in [0, 1]$ balances the contributions between static and dynamic anomaly.

Finally, an input $x$ is flagged as poisoned if its final anomaly score $S_{\text{final}}(x)$ exceeds a predetermined threshold $\tau$. We determine this threshold using the clean validation set $\mathcal{D}_{\text{valid}}$, targeting a false rejection rate of 5\%.

\subsection{Backdoor Purification based Unlearning}

To eliminate backdoor behaviors in the backdoored model, we propose a parameter-efficient unlearning approach based on LoRA fine-tuning. 
Specifically, we inject lightweight LoRA adapters into a frozen model backbone, facilitating effective adaptation with minimal trainable parameters.
However, the inherent information bottleneck associated with such parameter-efficient fine-tuning restricts its ability to eliminate deeply embedded backdoor knowledge~\cite{unlearningLLM}.

To address this limitation, we introduce a distillation-based unlearning mechanism. Specifically, we designate the original backdoored model as the teacher, with a copy initialized as the student. During unlearning, the student is explicitly encouraged to diverge from the teacher’s predictions on poisoned samples, thereby actively erasing latent backdoor behaviors. Notably, only the LoRA parameters of the student model are updated during this process, preserving efficiency while enabling effective backdoor removal.

A composite objective function $\mathcal{L}_{\text{total}}$ forms the foundation of our unlearning mechanism. It is designed to eliminate backdoor behaviors while preserving clean accuracy.
First, we introduce an unlearning loss $\mathcal{L}_{\text{unlearn}}$, which explicitly targets the removal of backdoor behavior from the backdoored model. Specifically, we employ the Kullback-Leibler (KL) to maximize divergence between the predictive distributions of the student and teacher models on poisoned samples $x' \in \mathcal{D}_p$:
\begin{equation}
    \mathcal{L}_{\text{unlearn}}=D_{\text{KL}}(\mathcal{M}_{\text{student}}(x')\parallel \mathcal{M}_{\text{teacher}}(x')), \text{for } x' \in \mathcal{D}_{p}.
\end{equation}

Second, to prevent degradation of clean accuracy during unlearning, we introduce a preservation loss $\mathcal{L}_{\text{preserve}}$. This loss uses standard Cross-Entropy (CE) to align the student model’s predictions with the ground-truth labels on clean samples $x \in \mathcal{D}_c$, effectively preserving clean knowledge:
\begin{equation}
    \mathcal{L}_{\text{preserve}} = \text{CE}(\mathcal{M}_{\text{student}}(x), y_{\text{true}}), \quad \text{for } x \in \mathcal{D}_{c}.
\end{equation}

The final training objective combines these two loss terms using tunable weights $\lambda_{\text{asr}}$ and $\lambda_{\text{acc}}$:
\begin{equation}
    \mathcal{L}_{\text{total}} = -\lambda_{\text{asr}} \cdot \mathcal{L}_{\text{unlearn}} + \lambda_{\text{acc}} \cdot \mathcal{L}_{\text{preserve}},
\end{equation}
where $\lambda_{\text{asr}}$ controls the degree of backdoor forgetting, while $\lambda_{\text{acc}}$ regulates the preservation of clean accuracy. Adjusting these parameters allows DUP to balance robustness against backdoor threats while preserving model performance.

\section{Experiments}
\subsection{Experimental Settings}

\noindent \textbf{Datasets} 
To comprehensively evaluate our method, we conduct experiments on three text classification datasets. For binary sentiment analysis, we use the \textbf{SST-2}~\cite{sst-2} and the \textbf{YELP}~\cite{YELP} dataset. For multi-class topic classification, we employ the \textbf{AG's News} dataset~\cite{AG}. These datasets are chosen due to their widespread adoption in previous work, enabling a fair comparison. 
The statistics of the datasets are in the Appendix~\ref{a:data}.

\noindent \textbf{Attack Setting}
We conduct experiments on four representative models to evaluate the effectiveness of our defense across diverse model scales and architectures. 
For PLMs, we use the encoder-only \textbf{BERT-base}~\cite{bert} and the encoder-decoder \textbf{BART-base}~\cite{bart}.
To assess performance on contemporary LLMs, we include two decoder-only models: \textbf{LLaMA-3.2-3B-Instruct}~\cite{llama32} and \textbf{Qwen-2.5-3B}~\cite{qwen2.5}. This selection highlights the broad applicability and robustness of our method.
We adhere to the hyperparameter settings established in previous work~\cite{synbkd,stylebkd} during training. Specifically, in line with~\cite{badacts}, we set the poisoning rate to 0.2 for generating poisoned training sets. All models are trained for 5 epochs using the AdamW optimizer~\cite{adamw} with an initial learning rate 2e-5 and a linear decay schedule. The top-$k$ parameter is set to $k = 3$.

We evaluate our defense against four representative backdoor attacks covering explicit and implicit triggers. For explicit-trigger attacks, we adopt: 1) \textbf{BadNets}~\cite{badnets}, which inserts a rare word (e.g., \texttt{"cf"}, \texttt{"mn"}, \texttt{"bb"}) as the trigger, and 2) \textbf{AddSent}~\cite{addsent}, which uses the fixed sentence \texttt{"I watch this 3D movie"} as the trigger. For implicit-trigger attacks, we use: 1) \textbf{Synbkd}~\cite{synbkd}, which adopt the syntactic template \texttt{"S(SBAR)(,)(NP)(VP)(.)"} as the trigger, and 2) \textbf{Stylebkd}~\cite{stylebkd}, which leverages the Bible style as the trigger. 
All experiments use the same computational environment, further specifications are provided in the Appendix~\ref{a:experiment}.

\noindent \textbf{Evaluation Metrics}
We evaluate detection performance using the \textbf{Area Under the Receiver Operating Characteristic (AUC)} as a threshold-independent metric, alongside the \textbf{False Acceptance Rate (FAR)} and the \textbf{False Rejection Rate (FRR)} for a more detailed analysis. For purification effectiveness, we report \textbf{Clean Accuracy (CACC)} to measure the utility, and \textbf{Attack Success Rate (ASR)} to assess the threat. The Appendix~\ref{a:metrics} contains detailed definitions of the metrics.

\begin{table*}[htb]
\centering
\setlength{\tabcolsep}{2.0mm}
\renewcommand{\arraystretch}{0.995}
\begin{tabular}{cccccccccccccc}
\toprule
\multirow{2}{*}{\textbf{Attack}} & \multirow{2}{*}{\textbf{Defense}} & \multicolumn{3}{c}{\textbf{BERT}} & \multicolumn{3}{c}{\textbf{BART}} & \multicolumn{3}{c}{\textbf{LLaMA 3B}} & \multicolumn{3}{c}{\textbf{Qwen 3B}} \\ \cmidrule(lr){3-5} \cmidrule(lr){6-8} \cmidrule(lr){9-11}  \cmidrule(lr){12-14}
&&AUC& FAR& FRR&AUC&FAR&FRR&AUC&FAR& FRR&AUC& FAR& FRR \\
\midrule
\multirow{4}{*}{BadNets}&STRIP&52.37&85.97&11.48&51.75&89.91&9.88&54.78&80.70&13.73&52.86&86.95&9.39\\
&DAN&90.97&40.68&5.49&83.64&71.93&\textbf{4.01}&62.21&88.82&5.60&64.66&87.61&6.10\\
&NAS&99.14&0.32&5.49&94.64&80.92&4.61&87.62&82.57&\textbf{5.33}&94.21&83.22&\textbf{4.89}\\
&MS&\textbf{100}&\textbf{0.00}&\textbf{5.44}&\textbf{99.27}&\textbf{0.22}&8.95&\textbf{94.39}&\textbf{32.13}&7.58&\textbf{98.29}&\textbf{0.44}&7.36\\ \cline{2-14}
\multirow{4}{*}{AddSent}&STRIP&53.95&87.17&11.53&50.44&91.23&7.74&51.63&90.57&8.68&54.38&84.54&11.53 \\
&DAN&57.96&95.61&4.94&74.38&93.09&\textbf{3.95}&59.31&89.69&\textbf{4.50}&55.42&98.03&5.77 \\
&NAS&99.45&\textbf{0.00}&5.99&85.86&89.91&4.56&93.30&86.84&4.72&96.75&7.02&\textbf{4.83} \\
&MS&\textbf{99.98}&\textbf{0.00}&\textbf{4.61}&\textbf{98.96}&\textbf{2.96}&6.26&\textbf{97.93}&\textbf{1.10}&7.63&\textbf{99.18}&\textbf{0.00}&6.53 \\ \cline{2-14}
\multirow{4}{*}{Stylebkd}&STRIP&53.99&88.82&9.77&53.68&85.75&10.38&52.66&91.67&8.07&53.03&88.60&11.26 \\
&DAN&79.75&64.25&5.99&93.37&44.41&\textbf{3.08}&77.82&66.34&5.16&80.83&61.40&4.83 \\
&NAS&81.91&60.20&\textbf{6.32}&99.71&\textbf{0.00}&3.95&97.65&9.32&\textbf{4.61}&98.03&\textbf{0.11}&\textbf{5.00} \\
&MS&\textbf{88.14}&\textbf{32.13}&6.43 &\textbf{99.77}&0.55&6.21&\textbf{99.44}&\textbf{0.00}&6.15&\textbf{98.71}&0.22&5.66 \\ \cline{2-14}
\multirow{4}{*}{Synbkd}&STRIP&50.97&93.64&\textbf{5.44}&50.46&95.94&4.94&51.00&88.16&11.53&51.85&88.93&9.77 \\
&DAN&77.19&81.69&6.15&86.62&70.29&\textbf{4.00}&70.73&90.13&5.71&67.46&95.50&5.49 \\
&NAS&72.77&91.67&5.71&90.28&86.40&4.39&91.50&78.18&\textbf{4.78}&91.09&85.64&\textbf{4.78} \\
&MS&\textbf{90.34}&\textbf{43.97}&6.15&\textbf{97.32}&\textbf{9.76}&7.36&\textbf{92.94}&\textbf{24.12}&8.46&\textbf{95.03}&\textbf{13.05}&7.03 \\
\bottomrule
\end{tabular}
\vspace{-0.5\intextsep}
\caption{Backdoor detection performance of MS and baselines on the SST-2 dataset. Metrics are reported in percentages (AUC, FAR, and FRR), and the best results are \textbf{highlighted in bold}.}
\label{t:detection_sst}
\end{table*}
\subsection{Backdoored Sample Detection}

\begin{table}[htb]
\centering
\renewcommand{\arraystretch}{0.95}
\begin{tabular}{ccccc|c}
\toprule
\textbf{Dataset} & \textbf{Metric} & \textbf{STRIP} & \textbf{DAN} & \textbf{NAS} & \textbf{MS} \\
\midrule
\multirow{3}{*}{SST-2} & AUC$\uparrow$&52.49&73.89&92.12&\textbf{96.85}\\
    & FAR$\downarrow$&88.66&77.47&52.65&\textbf{10.04}\\
     & FRR$\downarrow$&9.70&5.05&\textbf{5.00}&6.74\\ \cmidrule(l){2-6}
\multirow{3}{*}{YELP} & AUC$\uparrow$&53.20&86.06&97.94&\textbf{99.10}\\
       & FAR$\downarrow$&86.08&44.73&9.03&\textbf{2.60}\\
   & FRR$\downarrow$&10.55&7.14&\textbf{5.66}&6.33\\ \cmidrule(l){2-6}
\multirow{3}{*}{\begin{tabular}{c} AG's \\ News\end{tabular}}   & AUC$\uparrow$&53.03&92.20&93.67&\textbf{98.43}\\
& FAR$\downarrow$&81.14&23.04& 35.93&\textbf{6.39}\\
& FRR$\downarrow$&15.34&40.55& 4.96&\textbf{4.63}\\ 
\hline
\multirow{3}{*}{\textit{Average}} &AUC$\uparrow$& 52.91& 84.05&94.58 & \textbf{98.13}\\
& FAR$\downarrow$&85.29 &48.41 &32.54 & \textbf{6.34}\\ 
& FRR$\downarrow$& 11.86&17.58 &\textbf{5.21} &5.90 \\ 
\bottomrule
\end{tabular}
\caption{Average backdoor detection performance (in percentage) of our MS and baselines across four attack types (BadNets, AddSent, Stylebkd, and Synbkd) and four backdoored models (BERT, BART, LLaMA 3B, and Qwen 3B).}
\vspace{-0.75\intextsep}
\label{t:avg_detection}
\end{table}

\noindent \textbf{Overall Results} 
We compare MS against three backdoor defense methods: STRIP, DAN, and NAS. Table~\ref{t:avg_detection} summarizes the average detection performance of MS and the baselines. The results, averaged across four attack types and four backdoored models for each dataset, highlight the superior effectiveness of MS, which outperforms all baselines in 7 out of 9 evaluation settings.
\textbf{In terms of AUC and FAR, MS consistently surpasses all baselines across all datasets.}
Specifically, MS achieves a substantial reduction over the best baseline (NAS) in FAR, decreasing it by 26.20\% on average.
Meanwhile, MS maintains a competitive FRR at a low average of 6.34\%.


Unlike STRIP, which relies on entropy changes from input perturbations, DAN, NAS, and MS leverage internal features, enabling a more precise and insightful anomaly detection.
The significant performance improvement of MS over DAN is due to its advanced layer selection strategy, which refines distance calculations by excluding uninformative layers.
Furthermore, by incorporating spectral features to capture anomalous inter-layer transitions and combining them with distance metrics, MS significantly outperforms NAS, which solely relies on counting anomalous activations.
Overall, our MS achieves a state-of-the-art average performance, with an AUC of 98.13\%, and maintains low average FAR and FRR values of 6.34\% and 5.90\%, respectively.

Table~\ref{t:detection_sst} provides detailed results on the SST-2 dataset. The Appendix~\ref{a:detection} presents detection results for the other two datasets. These results indicate that the efficacy of baseline methods strongly depends on the underlying model architecture.
For example, DAN performs well against BadNets on BERT (90.97\% AUC), but its performance drops significantly on LLaMA 3B (62.21\% AUC), which highlights its limited generalizability.
Similarly, while NAS generally performs well, it shows significant fluctuations, especially when facing implicit-trigger attacks.
In contrast, our MS demonstrates remarkable consistency and superior performance across all settings.
Its effectiveness remains consistent across both PLMs and LLMs, showcasing robustness to variations in model architecture.
Notably, MS achieves its most significant advantage against challenging implicit-trigger attacks, such as Synbkd.
Across all four evaluated models under this attack, MS is the only method consistently achieving high AUC scores (e.g., 90.34\% on BERT) and low FAR values (e.g., 13.05\% on Qwen 3B), whereas the baselines perform worse.
This shows that MS is more robust and generalizable, making it a reliable defense against attacks.

\begin{table}[htb]
\renewcommand{\arraystretch}{0.975}
\begin{tabular}{ccccc}
\toprule
\textbf{Models}& \textbf{Setting}& \textbf{BERT} & \textbf{BART} &\textbf{LLaMA 3B} \\
\midrule
\multirow{4}{*}{BadNets}  & \textit{first half}  &88.10&58.53&51.31\\
        & \textit{last half} &99.84&81.15&90.16\\ 
        & \textit{all} &99.56&74.73&80.01\\
        & \textit{top-k}&\textbf{100}&\textbf{99.27}&\textbf{94.39}\\ 
        \cline{2-5}
\multirow{4}{*}{AddSent}  & \textit{first half} &99.15&53.05&68.89\\
        & \textit{last half} &99.92&97.41&96.33\\ 
        &\textit{all}  &99.80&87.24&91.49\\
        & \textit{top-k} &\textbf{99.98}&\textbf{98.96}&\textbf{97.93}\\ 
        \cline{2-5}
\multirow{4}{*}{Stylebkd}  & \textit{first half} &75.91&85.24&97.83\\
        & \textit{last half}&85.29&99.49&98.94\\ 
        & \textit{all}  &84.84&98.12&98.50\\
        & \textit{top-k} &\textbf{88.14}&\textbf{99.78}&\textbf{99.44}\\ 
        \cline{2-5}
\multirow{4}{*}{Synbkd}  &\textit{first half} &52.87&68.32&88.62\\
        &\textit{last half}&83.64&96.01&98.94\\ 
        & \textit{all}  &76.54&89.24&92.51\\
        & \textit{top-k} &\textbf{90.34}&\textbf{97.32}&\textbf{92.94}\\
\bottomrule
\end{tabular}
\caption{Backdoor detection performance (AUC in percentage) of MS with different layer selection strategy on SST-2.}
\vspace{-0.5\intextsep}
\label{t:md}
\end{table}

\noindent \textbf{Ablation Experiments}
To validate the effectiveness of the \textit{top-k} layer selection strategy, we compare the proposed MS, which dynamically selects the most informative layers based on the CH score, against three baselines: using only the first half of layers (\textit{first half}), only the last half of layers (\textit{last half}), and all available layers (\textit{all}). From Table~\ref{t:md}, we observe a clear pattern where using deeper layers (i.e., \textit{last half}) consistently yields better results than using shallower layers (\textit{first half}). This observation indicates that deeper layers possess more discriminative features for backdoor detection. 
However, naively including all layers often leads to inferior performance, likely due to the noisy or irrelevant features from shallower layers.
By adaptively identifying and focusing on the most discriminative layers, \textbf{the proposed \textit{top-k} strategy consistently achieves superior performance}, effectively mitigating this issue.
The impact of the hyperparameter $k$ is analyzed in the Appendix~\ref{a:layers}.

\begin{table}[htb]
\renewcommand{\arraystretch}{0.975}
\begin{tabular}{ccccc}
\toprule
\textbf{Models}& \textbf{Setting}& \textbf{BERT} & \textbf{BART} &\textbf{LLaMA 3B} \\
\midrule
\multirow{2}{*}{BadNets}  & \textit{w/o ss} &99.90&85.71&\textbf{95.61}\\
                        & \textit{w/ ss}   &\textbf{100}&\textbf{99.27}&94.39\\
                        \cline{2-5}
\multirow{2}{*}{AddSent}  &\textit{w/o ss} &99.97&98.69&\textbf{98.21}\\
                        & \textit{w/ ss}  &\textbf{99.98}&\textbf{98.96}&97.93\\
                        \cline{2-5}
\multirow{2}{*}{Stylebkd } & \textit{w/o ss} &88.09&99.76&99.42\\
& \textit{w/ ss} & \textbf{88.14} & \textbf{99.78} & \textbf{99.44} \\ \cline{2-5}
\multirow{2}{*}{Synbkd } & \textit{w/o ss} &89.30&97.28&91.83\\
                        & \textit{w/ ss}   &\textbf{90.34}&\textbf{97.32}&\textbf{92.94}\\
\bottomrule
\end{tabular}
\vspace{-0.25\intextsep}
\caption{Backdoor detection performance (AUC in percentage) of MS with and without spectral signatures (\textit{w/ ss} and \textit{w/o ss}) across different models on the SST-2 dataset.}
\label{t:ss}
\end{table}

Furthermore, we conduct an ablation study to evaluate the effectiveness of Spectral Signatures (\textit{ss}).
We compare our complete method (\textit{w/ ss}) against a variant that relies solely uses distance without the spectral (\textit{w/o ss}).
As shown in Table~\ref{t:ss}, the results demonstrate that \textbf{incorporating spectral achieves superior performance across most scenarios (10 out of 12}.
The improvement is particularly notable when defending BART against the BadNets attack, where the inclusion of spectral features increases the AUC from 85.71\% to 99.27\%.
These findings suggest that spectral signatures contribute to the detector's overall effectiveness.
The impact of the fusion weight $\alpha$ is analyzed in the Appendix~\ref{a:alpha}. Moreover, we conduct an ablation study on the impact of the number of clean samples on detection performance; detailed results are provided in Appendix~\ref{a:cleandata}.

\subsection{Backdoored Model Purification}

\begin{table*}[htb]
    \centering
    \renewcommand{\arraystretch}{0.985}
    \begin{tabular}{cccccccccc}
        \toprule
        \multirow{2}{*}{\textbf{Attack}} & \multirow{2}{*}{\textbf{Defense}}  & \multicolumn{2}{c}{\textbf{BERT}} & \multicolumn{2}{c}{\textbf{BART}} & \multicolumn{2}{c}{\textbf{LLaMA 3B}}& \multicolumn{2}{c}{\textbf{Qwen 3B}}\\ \cmidrule(lr){3-4} \cmidrule(lr){5-6} \cmidrule(lr){7-8} \cmidrule(lr){9-10}
         & & CACC$\uparrow$ & ASR$\downarrow$ & CACC$\uparrow$ & ASR$\downarrow$  & CACC$\uparrow$ & ASR$\downarrow$  & CACC$\uparrow$ & ASR$\downarrow$  \\
         \midrule
         \multirow{4}{*}{BadNets} & ONION & 81.27& 38.93&85.06&30.26 &86.49 &23.25 &85.34 &26.32\\
         & BadActs &82.40&37.83&-&-&-&-&-&-\\
         & TG &85.28&32.46&83.25&34.54&89.07&48.57&61.07&90.46\\
         & DUP & \textbf{88.96} &\textbf{1.64} &\textbf{91.27} &\textbf{3.84}&\textbf{91.98} & \textbf{1.43}& \textbf{91.21}&\textbf{0.22}\\ \cline{2-10}
        \multirow{4}{*}{AddSent} & ONION &85.56 &93.75&89.07 &94.85 &87.70 &83.99 & \textbf{86.27}&73.46\\ 
         & BadActs &71.35&53.86&-&-&-&-&-&-\\
         & TG &82.87&45.94&87.26&23.36&\textbf{92.97}&17.76&65.13&65.46\\
         & DUP & \textbf{90.17}& \textbf{0.22}&\textbf{90.94}&\textbf{2.19}& 90.06& \textbf{0.00}&83.53 &\textbf{0.00}\\ \cline{2-10}
        \multirow{4}{*}{Stylebkd} & ONION &85.28 &83.63& 87.42&99.78 &75.89 &99.89  & 75.73&99.78\\
         & BadActs &76.88&45.55&-&-&-&-&-&-\\
         & TG &87.81&75.00&87.86& 92.21&\textbf{93.19}& 96.82& \textbf{91.49}&93.09\\
         & DUP &\textbf{90.06}&\textbf{5.81}& \textbf{91.71}&\textbf{3.18} &90.61&\textbf{0.00}&87.26 &\textbf{0.00}\\ \cline{2-10}
        \multirow{4}{*}{Synbkd} & ONION &85.50&90.68&86.11 &96.16 &76.22  &98.68   &82.92 &95.50\\ 
         & BadActs &78.59&42.08&-&-&-&-&-&-\\
         & TG &87.86&43.86&87.75&49.23&\textbf{92.97}&55.92&77.38&77.30\\
         & DUP &\textbf{89.13}&\textbf{5.48}& \textbf{90.55}& \textbf{1.21}&90.23 &\textbf{0.66} & \textbf{83.96}&\textbf{0.00}\\
         \bottomrule
    \end{tabular}
    \vspace{-0.5\intextsep}
    \caption{Comparison of purification performance (CACC and ASR in percentage) between DUP and baseline defenses across four attack types and four model architectures on the SST-2 dataset. BadActs is implemented only for BERT.} 
    \label{t:puri_plm}
\end{table*}

\noindent \textbf{Overall Results} 
This part compares DUP with three baselines: ONION, BadActs, and TG.
As shown in Table \ref{t:puri_plm}, our DUP demonstrates superior performance over all baseline defenses across the four models and four distinct backdoor attacks on the SST-2 dataset. 
\textbf{DUP achieves the highest CACC in the majority of settings (11 out of 16), and in terms of ASR, it achieves the lowest ASR in all settings.}
For example, against the AddSent attack, DUP reduces the ASR to 0.22\% on BERT and 0.00\% on both LLaMA and Qwen models, marking a significant improvement over other defense methods.
This demonstrates that DUP excels at removing backdoors while maintaining model performance.

\begin{figure}[htb]
    \centering
    \includegraphics[width=0.8\linewidth]{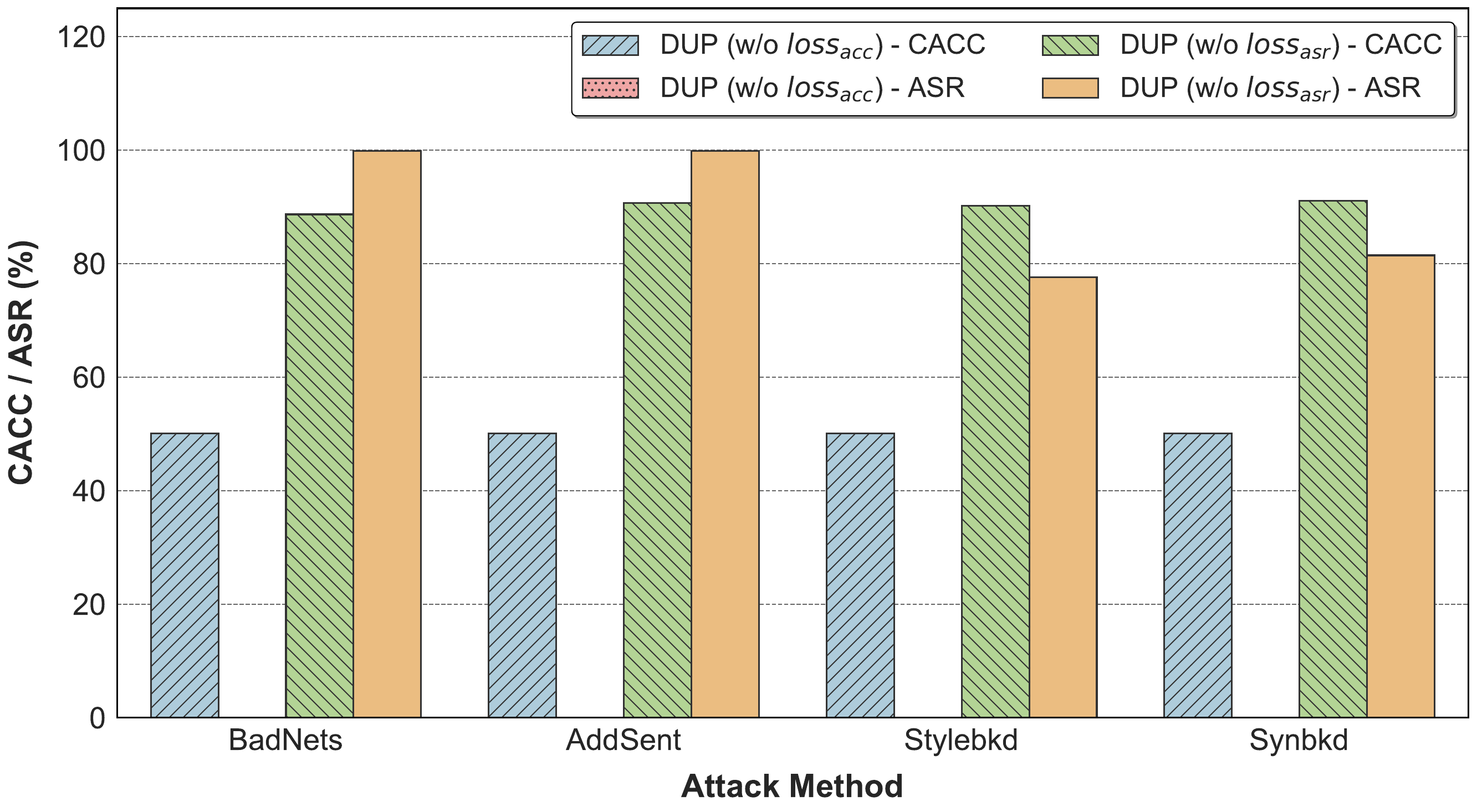}
    \vspace{-0.5\intextsep}
    \caption{The impact of different loss components on the purification performance of DUP on BERT (SST-2).}
    \label{f:loss}
\end{figure}

Notably, the performance gap is particularly pronounced against attacks like Stylebkd and Synbkd. Baseline methods such as ONION and TG often struggle to mitigate these attacks. They typically exhibit ASR exceeding 75\%. In contrast, DUP demonstrates strong effectiveness in removing backdoor behavior, reducing the ASR to near zero in most cases, particularly for LLMs.
This demonstrates the DUP's adaptability in handling various backdoor threats, from basic trigger insertions to more advanced attacks.

\noindent \textbf{Ablation Experiments}
We conduct an ablation study to verify the efficacy of the two components of our objective function. As shown in Figure~\ref{f:loss}, removing the preservation loss ($\lambda_{acc}=0$) results in a significant degradation in CACC, rendering the model unusable despite completely eliminating the backdoor (ASR=0). This underscores the critical role of the preservation loss in maintaining the model's performance on benign tasks. Conversely, when the unlearning loss is removed ($\lambda_{asr}=0$), the model's CACC remains high, but the ASR is largely unaffected. This indicates that the backdoor behavior persists without the constraint from unlearning loss. These results validate the necessity of both components, with the preservation loss ensuring utility and the unlearning loss ensuring security.

\begin{figure}[t]
\centering
\begin{subfigure}[b]{0.48\linewidth}
    \centering
    \includegraphics[width=\linewidth]{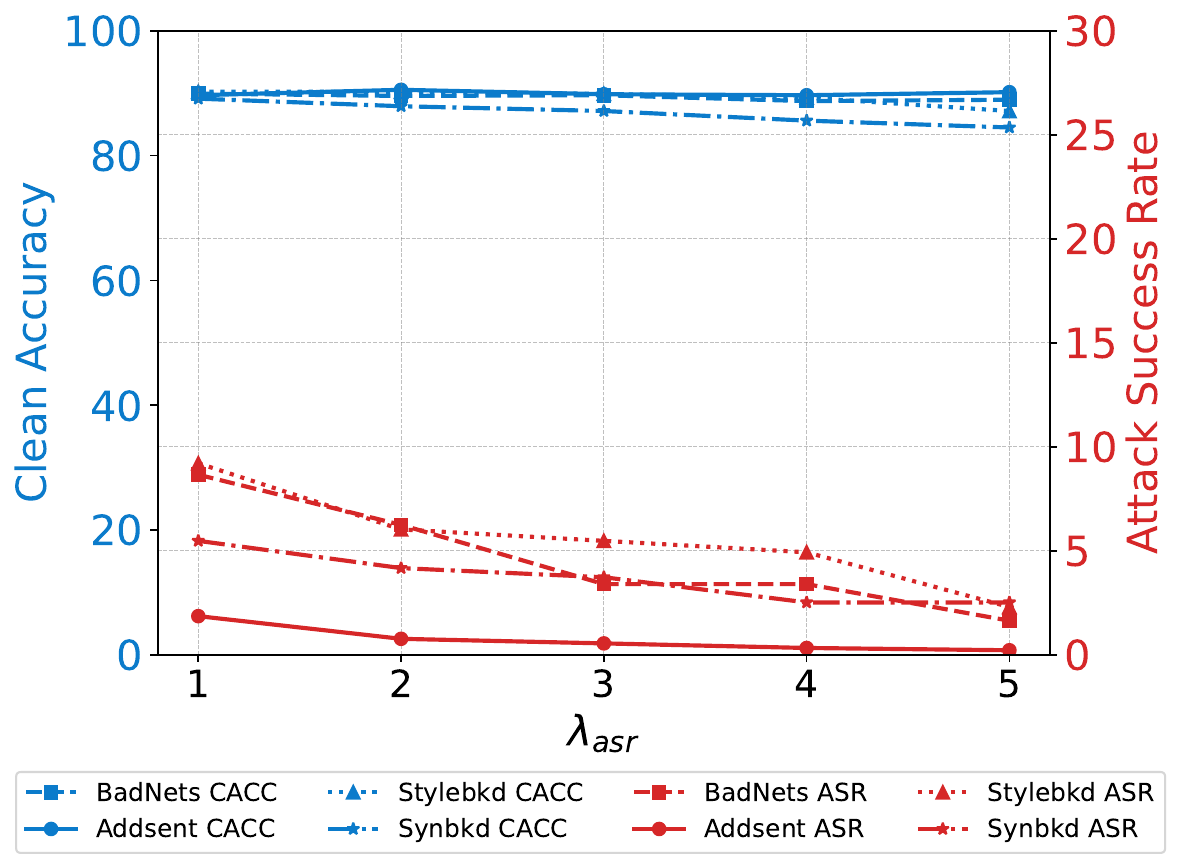}
\end{subfigure}
\hfill
\begin{subfigure}[b]{0.48\linewidth}
    \centering
\includegraphics[width=\linewidth]{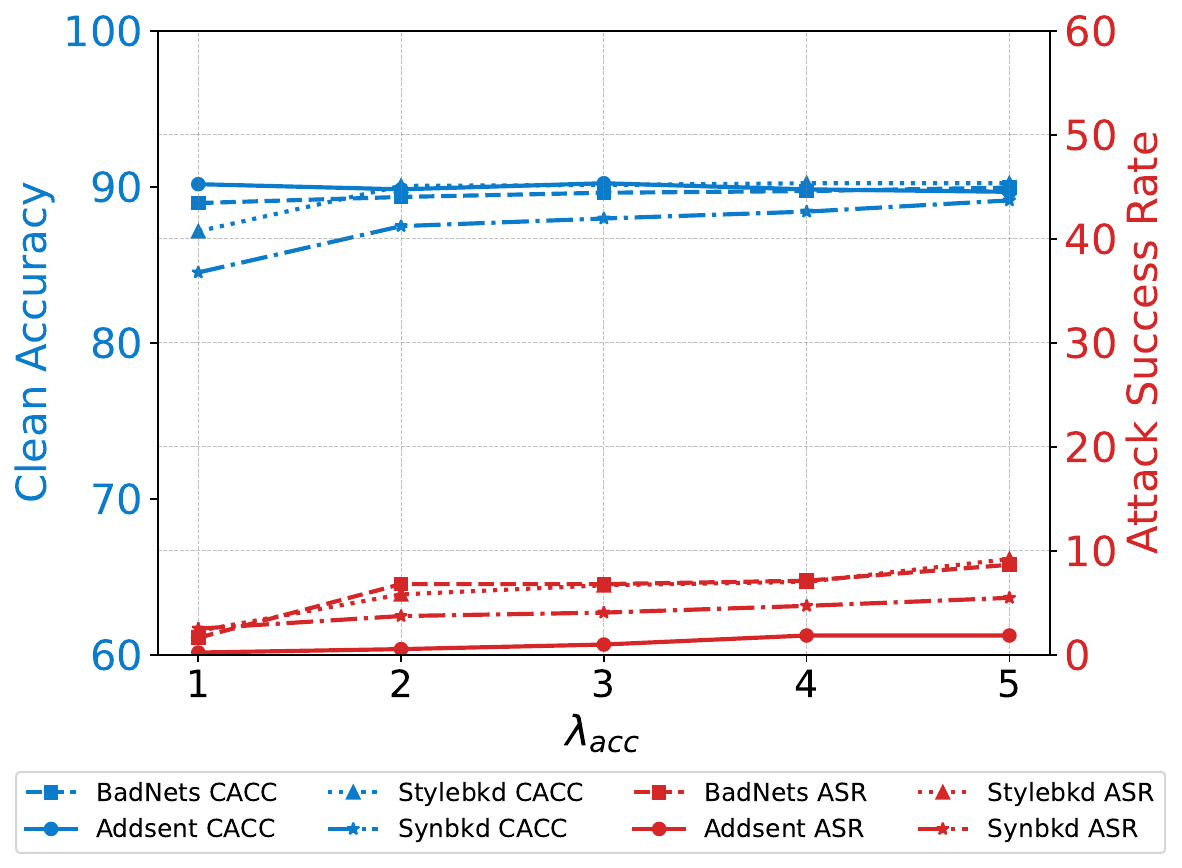}
\end{subfigure}
\vspace{-0.5\intextsep}
\caption{Impact of $\lambda_{asr}$ and $\lambda_{acc}$ on purification performance of DUP on BERT (SST-2). Left: Varying $\lambda_{asr}$ with fixed $\lambda_{acc}$. Right: Varying $\lambda_{acc}$ with fixed $\lambda_{asr}$.}
\vspace{-0.5\intextsep}
\label{f:puri_ablation}
\end{figure}

To investigate the impact of the hyperparameters $\lambda_{acc}$ and $\lambda_{asr}$ on the performance of our DUP, we conduct experiments by fixing one while varying the other in the range from 1 to 5. 
The results are shown in Figure~\ref{f:puri_ablation}, where the left subfigure varies $\lambda_{asr}$ with fixed $\lambda_{acc}$, and the right vice versa. From the left subfigure, increasing $\lambda_{asr}$ consistently reduces the ASR across different attack methods. This indicates that \textbf{the knowledge-distillation-based unlearning loss effectively removes backdoor behaviors from the student model.}
On the other hand, increasing $\lambda_{acc}$ leads to improvement in CACC, suggesting that \textbf{the cross-entropy-based preservation loss enables the student model to maintain its normal performance.}
As illustrated in both subfigures, our DUP demonstrates robust performance, with CACC and ASR remaining relatively stable throughout the variation of both hyperparameters. This highlights the reliability of DUP in balancing attack mitigation and clean accuracy preservation across different configurations.

\subsection{Robustness to Adaptive Attacks}
We evaluate the robustness of DUP against adaptive attacks by employing feature-level regularization. Building on previous work in computer vision~\cite{Zhao0XDWL22,ZhongQZ22}, we regularize poisoned samples to match the latent representations of clean samples. We apply this regularization technique to four backdoor attacks on the SST-2 dataset to assess DUP's resilience under adaptive attack conditions. \textbf{As shown in Table~\ref{t:reg}, DUP demonstrates only a slight decline in performance, highlighting its robustness to adaptive attacks.} Despite the adaptive attack reducing the distance between poisoned and clean features, spectral discrepancies continue to offer valuable signals for detecting poisoned samples. We provide detailed implementation of the adaptive attack in Appendix~\ref{a:adaptive}.

\begin{table}[htb]
    \centering
    \renewcommand{\arraystretch}{0.7}
    \begin{tabular}{ccccc}
    \toprule
        \textbf{Attack} & \textbf{Setting} & \textbf{CACC}$\uparrow$ & \textbf{ASR}$\downarrow$  \\
        \midrule
        \multirow{2}{*}{BadNets} & \textit{w/o reg} & 88.96  & 1.64  \\
         &\textit{ w/ reg}&  89.62& 3.51 \\ \cline{2-4}
         \multirow{2}{*}{AddSent} & \textit{w/o reg} & 90.17  &  0.22 \\
         & \textit{w/ reg}& 89.62 & 0  \\ \cline{2-4}
         \multirow{2}{*}{Stylebkd} & \textit{w/o reg} & 90.06  &  5.81 \\
         & \textit{w/ reg} & 90.94 &  12.28 \\ \cline{2-4}
         \multirow{2}{*}{Synbkd} & \textit{w/o reg} & 89.13  &  5.48 \\
         & \textit{w/ reg} & 86.49 & 8.44  \\
         \bottomrule
    \end{tabular}
    \caption{Purification performance (in percentage) of DUP with and without feature-level regularization (\textit{reg}) adaptive attacks on BERT (SST-2).}
    \vspace{-0.75\intextsep}
    \label{t:reg}
\end{table}

\section{Related Works}

Existing backdoor defense methods can be broadly categorized into three directions: 
(1) \textbf{Backdoor suppression}, which mitigates the influence of backdoor behaviors by isolating backdoor functionality~\cite{st} or leveraging ensemble-based strategies~\cite{TG};
(2) \textbf{Backdoor detection}, which operates at the input level by applying perturbations to observe variations in entropy or perplexity~\cite{onion,rap,strip}, or at the feature level by analyzing inconsistencies in the model’s internal activations~\cite{dan,cube,badacts}.
(3) \textbf{Backdoor purification}, which aims to eliminate backdoors from the backdoored models using techniques such as token unlearning~\cite{btu}, activation clipping\cite{badacts}, and knowledge distillation~\cite{unlearningLLM}.
The Appendix~\ref{a:relatedwork} includes detailed discussions of related work.
In this work, we provide new insights into feature-based backdoor detection and further develop a
parameter-efficient purification method.

\section{Conclusion}
In this paper, we propose DUP (Detection-guided Unlearning for Purification), a unified framework that integrates feature-space backdoor detection with parameter-efficient unlearning techniques to defend backdoor attacks in language models. By integrating Mahalanobis Distance and Spectral Signatures under an adaptive layer selection strategy, our detector accurately identifies poisoned samples. Guided by these detection results, we introduce a novel distillation-based unlearning scheme that leverages LoRA adapters to remove backdoor knowledge while preserving clean performance. 
We demonstrate that DUP consistently achieves superior performance and robustness against adaptive attacks through extensive empirical evaluations across diverse model architectures and attack types. 
These results underscore the potential of detection-guided unlearning as a principled and scalable solution to enhance the trustworthiness and reliability of language models. 
Future work may investigate its application to multimodal models.

\bibliography{main}

\appendix

\section{More Related Work}
\label{a:relatedwork}
\noindent \paragraph{Backdoor Attack}
Backdoor attacks in natural language processing (NLP) are commonly categorized based on the granularity and visibility of the trigger~\cite{lin2024backtime,wen2025investigating,hu2025syntactic,zhao2025survey}.
Early methods focus on character-level triggers, such as inserting or substituting rare characters~\cite{badnl,li2021hidden}.
An alternative class of attacks utilizes word-level triggers, where specific keywords or phrases are either inserted or substituted into input texts to induce targeted behavior~\cite{kurita2020weight,qi2021turn,bite}. These triggers are often chosen for their low natural frequency or semantic neutrality to avoid detection.
In sentence-level attacks, predefined emotionally neutral sentences are appended to the input to activate the backdoor~\cite{addsent}, making them harder to detect while maintaining grammatical fluency.
To further improve stealthiness, some approaches have introduced implicit triggers, which embed the trigger in global textual properties rather than discrete character or word insertions. Notably, style-based attacks manipulate text style~\cite{stylebkd}, while syntactic backdoors utilize specific grammatical templates to encode the trigger~\cite{synbkd}.
As backdoor techniques evolve toward more covert, generalizable, and model-agnostic designs, they increasingly evade input-level perturbation detection. This poses serious challenges to defense mechanisms that rely on input-level cues or coarse-grained feature representations.

\noindent \paragraph{Backdoor Defense}

Various defense strategies have been developed to mitigate backdoor threats in NLP models.
STRIP~\cite{strip} is an input-level detection method that identifies poisoned samples by measuring changes in prediction entropy under input perturbations. A low entropy variance indicates backdoor activation.
DAN~\cite{dan} focuses on feature-level detection, leveraging the observation that poisoned and clean samples exhibit distinct intermediate activations. It computes anomaly scores based on distances in feature space to identify malicious inputs.
BadActs~\cite{badacts} proposes a model-level purification strategy by learning clean activation intervals and clipping anomalous activations of suspicious inputs back into these intervals, effectively suppressing backdoor effects.
TextGuard (TG)~\cite{TG} takes a different approach by integrating ensemble learning to reduce the model's reliance on specific triggers, thereby diluting the backdoor effect without explicit detection. While these methods have shown promising results, they often suffer from high inference overhead (e.g., STRIP), limited generalizability to implicit triggers (e.g., DAN), or degradation of model utility (e.g., BadActs, TG).

\section{Dataset Statistics}
\label{a:data}
Table~\ref{t:data_stats} summarizes the statistics of the benchmark datasets used in our experiments, including task type, class distribution, data splits, and average text length. We conducted experiments on SST-2 and YELP for sentiment classification, and AG's News for topic classification. The target class used in backdoor attacks is marked in \textbf{bold}:
\begin{itemize}
    \item SST-2~\cite{sst-2} is a binary sentiment analysis dataset derived from the Stanford Sentiment Treebank movie reviews. It contains short, well-formed sentences. We use the \textbf{Positive} class as the target for attack.

    \item YELP~\cite{YELP} is another binary sentiment dataset consisting of user reviews collected from the YELP platform. Compared to SST-2, it contains longer and more diverse texts. The \textbf{Positive} class is selected as the attack target.

    \item AG's News~\cite{AG} is a multi-class topic classification dataset composed of news articles categorized into four classes: World, Sports, Business, and Tech. Each article typically contains a title and a short description. We designate the \textbf{World} category as the attack target in our experiments. 
\end{itemize}
These datasets span binary and multi-class classification tasks, offering a diverse and challenging benchmark for evaluating backdoor defenses across text lengths, domains, and label diversity.

\begin{table*}[ht]
    \centering
    \begin{tabular}{ccccc}
         \toprule
         \textbf{Datasets} & \textbf{Task} & \textbf{Class} & \textbf{Split (Train/Dev/Test)}  & \textbf{Avg. Len.} \\
         \midrule
         SST-2 & Sentiment Analysis & \textbf{Positive}/Negative & 6.9K:0.8K:1.8K & 19.21 \\
         YELP & Sentiment Analysis & \textbf{Positive}/Negative & 14K:3K:3K & 29.71 \\
         AG's News& News Topic Classification & \textbf{World}/Sports/Business/SciTech& 108K:12K:7.6K& 31.06 \\
         \bottomrule
    \end{tabular}
    \caption{Statistics of the benchmark datasets used in our experiments. The target class is marked in \textbf{bold}.}
    \label{t:data_stats}
\end{table*}

\section{Algorithmic Description}
\label{a:algo}
Algorithm~\ref{alg:msdefender} outlines our DUP framework, which purifies backdoored models while preserving clean-task performance. The method consists of two stages, which are poisoned sample detection and targeted unlearning. First, we identify informative layers by computing the Calinski–Harabasz score on a calibration set $\mathcal{D}_{calib}$. Mahalanobis distances are computed for each selected layer as a distance-based anomaly score. In parallel, we extract inter-layer spectral features via singular value decomposition to capture inconsistencies caused by backdoor triggers. These two scores are normalized and combined to form a final detection score. The threshold $\tau$ is set by fixing the false rejection rate at 5\%, and samples are classified into poisoned or clean sets accordingly. Next, we fine-tune a student model with LoRA adapters using a dual-loss objective that includes KL divergence to forget backdoor behavior and cross-entropy loss to retain clean accuracy. The total loss is formulated as a weighted combination of the two objectives, with weights $\lambda_{\text{asr}}$ and $\lambda_{\text{acc}}$.

\begin{algorithm}[ht]
\caption{Detection-guided Unlearning for Purification}
\label{alg:msdefender}
\begin{algorithmic}[1]
\REQUIRE Backdoored model (teacher) $\mathcal{M}$, calibration set $\mathcal{D}_{calib}$, validation set $\mathcal{D}_{\text{valid}}$, test input $x$, loss weights $\lambda_{\text{asr}}, \lambda_{\text{acc}}$, FRR=5\%;
\ENSURE Predicted labels; purified model (student) $\mathcal{M}_s$;

\STATE Select $k$ layers by Calinski Harabasz score on $\mathcal{D}_{calib}$;
\FOR{each selected layer $i$}
    \STATE Compute class-agnostic centroids $c_i$ and  shared shrunk covariance $\boldsymbol{\Sigma}_i$;
\ENDFOR

\STATE Extract features $f_i(x)$ from each layer;
\STATE $M_i(x) \leftarrow \sqrt{(f_i(x) - c_i)^T \boldsymbol{\Sigma}_i^{-1} (f_i(x) - c_i)}$;
\STATE $S_{\text{MD}}(x) \leftarrow$ mean of $\{M_i(x)\}_{i=1}^k$;

\STATE Stack hidden states $\mathbf{H}(x) = [f_1(x), \dots, f_L(x)]^\top$;
\STATE Calculate the inter-layer difference matrix $\boldsymbol{\Delta}(x)$;
\STATE Perform SVD: $\boldsymbol{\Delta}(x) = U \Sigma V^\top$;
\STATE $S_{\text{SS}}(x) \leftarrow s_1 / \sum_j s_j$;
\STATE Normalize both scores to get $\hat{S}_{\text{MD}}(x)$ and $\hat{S}_{\text{SS}}(x)$;
\STATE $S_{\text{final}}(x) = \alpha \cdot \hat{S}_{\text{MD}}(x) + (1 - \alpha) \cdot \hat{S}_{\text{SS}}(x)$;
\STATE Compute threshold $\tau$ corresponding to 5\% FRR;
\IF{$S_{\text{final}}(x) > \tau$}
    \STATE $\texttt{labels} \leftarrow \texttt{poisoned}$; add $x$ to $\mathcal{D}_p$;
\ELSE
    \STATE $\texttt{labels} \leftarrow \texttt{clean}$; add $x$ to $\mathcal{D}_c$;
\ENDIF

\STATE Copy $\mathcal{M}$ as $\mathcal{M}_{\text{s}}$ and insert LoRA adapters;
\FOR{each batch $(x, y)$ from $\mathcal{D}_p \cup \mathcal{D}_c$}
    \STATE $\mathcal{L}_{\text{unlearn}} = D_{\text{KL}}(\mathcal{M}_s(x) \parallel \mathcal{M}(x))$ on $\mathcal{D}_p$;
    \STATE $\mathcal{L}_{\text{preserve}} = \text{CE}(\mathcal{M}_s(x), y)$ on $\mathcal{D}_c$;
    \STATE $\mathcal{L}_{\text{total}} = -\lambda_{\text{asr}} \cdot \mathcal{L}_{\text{unlearn}} + \lambda_{\text{acc}} \cdot \mathcal{L}_{\text{preserve}}$;
\ENDFOR

\RETURN $\texttt{labels}$, $\mathcal{M}_{\text{s}}$
\end{algorithmic}
\end{algorithm}

\section{Experimental Settings}
\label{a:experiment}
We reproduced six representative baseline defense methods: STRIP~\cite{strip}, DAN~\cite{dan}, NAS~\cite{badacts}, ONION~\cite{onion}, BadActs~\cite{badacts}, and TG~\cite{TG}. These methods span different defense strategies, including backdoor input detection, model-level purification, and backdoor suppression.

We adopted publicly available implementations for all baseline methods and followed standardized configurations to ensure fair comparison. For BadActs, we set the threshold margin $\delta$ to 3 and fixed the FRR at 5\%. ONION was configured with a perplexity threshold of 0. For STRIP, we used five repetitions, a word swap ratio of 0.5, and turned off the use of opposite-label sets. DAN was similarly evaluated under an FRR of 5\%. TG was trained for five epochs, with the number of ensemble groups set to 9 for PLMs and 3 for LLMs due to GPU memory constraints. For our MS, we selected the top-$k = 3$ layers for Mahalanobis-based scoring and used all layers for spectral scoring. The final anomaly score was computed using a weighted fusion with $\alpha = 0.9$, and the detection threshold was calibrated to maintain an FRR of 5\%.
We fixed all random seeds to 2025 to ensure reproducibility.

All baselines were executed under consistent computational settings for fair comparison. To efficiently fine-tune backdoored models, we employed LoRA adapters via Hugging Face's PEFT library, with configuration parameters: rank $r=32$, $\alpha=64$, dropout rate 0.1, and no bias adaptation. In memory-constrained scenarios, such as full-parameter fine-tuning of 3B-scale models during backdoor implantation, we utilized Hugging Face Accelerate and DeepSpeed ZeRO Stage 1 to offload optimizer states to the CPU. Training was conducted using \texttt{bf16} precision and a gradient accumulation step size of 8.

All experiments were implemented using the PyTorch~\cite{torch}, OpenBackdoor~\cite{cube}, PEFT~\cite{peft}, and Accelerate~\cite{accelerate} libraries.
They were conducted on a machine equipped with a single NVIDIA RTX 4090 GPU (24 GB VRAM) and a 32 vCPU Intel(R) Xeon(R) Gold 6430 processor. Due to memory constraints, the TG baseline was executed using two RTX 4090 GPUs.

\section{Evaluation Metrics}
\label{a:metrics}
We use the Area Under the Receiver Operating Characteristic (AUC) to evaluate our detection method as a primary, threshold-independent measure of its overall efficacy. Additionally, we report the False Acceptance Rate (FAR), the proportion of clean samples incorrectly flagged as poisoned, and the False Rejection Rate (FRR), the proportion of poisoned samples missed by the detector. A robust defense must strike a balance between keeping FAR low to preserve clean input utility and maintaining FRR low to ensure security against backdoor attacks.
We assess the effectiveness of our purification method and its impact on model utility through two key metrics, Clean Accuracy (CACC) and Attack Success Rate (ASR). CACC measures the purified model's accuracy on clean samples, which quantifies its performance preservation. Conversely, ASR is the percentage of poisoned samples still misclassified as the target label. 

\section{Effectiveness of Backdoor Attacks Across Models and Datasets}
\label{a:attack}
Table~\ref{t:attack_results} presents the ASR and CACC of four representative backdoor attacks (BadNets, AddSent, Stylebkd, and Synbkd) across different models and datasets. All attacks consistently achieve high ASR while preserving high CACC, demonstrating their robustness across both PLMs and LLMs. The strength and stealth of these attacks present substantial challenges to current defense methods.

\begin{table*}[ht]
\centering
\begin{tabular}{ccccccccccc}
\toprule
\multirow{2}{*}{Dataset} & \multirow{2}{*}{Attack} & \multicolumn{2}{c}{BERT-base} & \multicolumn{2}{c}{BART-base} & \multicolumn{2}{c}{LLaMA 3B} & \multicolumn{2}{c}{Qwen 3B} \\  \cmidrule(lr){3-4} \cmidrule(lr){5-6} \cmidrule(lr){7-8}   \cmidrule(lr){9-10} 
              &                & CACC  & ASR   & CACC  & ASR   & CACC  & ASR   & CACC   & ASR  \\ \midrule
\multirow{4}{*}{SST-2}&BadNets &90.72\%&100\%  &90.12\%&100\%  &94.95\%&99.89\%&94.23\%&100\%\\
                      &AddSent &90.28\%&100\%  &93.08\%&100\%  &94.12\%&100\%  &95.66\%&100\%\\
                      &Stylebkd&89.90\%&81.69\%&92.97\%&99.78\%&94.84\%&99.89\%&95.28\%&99.56\%\\
                      &Synbkd  &90.39\%&89.69\%&91.93\%&96.71\%&91.76\%&99.34\%&93.85\%&96.49\%\\ \hline
\multirow{4}{*}{YELP} &BadNets &95.54\%&100\%  &96.04\%&99.93\%&97.10\%&99.80\%&96.63\%&100\%\\
                      &AddSent &95.44\%&100\%  &96.73\%&100\%  &97.23\%&100\%&95.84\%&100\%\\
                      &Stylebkd&94.60\%&93.60\%&95.40\%&100\%  &97.33\%&100\%&97.13\%&100\%\\
                      &Synbkd  &95.44\%&99.67\% &95.87\%&100\% &96.70\%&100\%&97.23\%&100\%\\ \hline
\multirow{4}{*}{AG's News}&BadNets&94.43\%&100\%&94.90\%   &100\% &95.07\%&100\%&94.53\%&99.88\%\\
                        &AddSent &94.33\%&100\%&94.65\%  &100\% &95.04\%&100\%&94.65\%&100\%\\
                        &Stylebkd&94.24\%&94.05\%&94.57\%&95.86\%&94.71\%&95.56\%&94.67\%&96.16\%\\
                        &Synbkd  & 94.17\%&99.75\% &94.79\% &99.91\% & 94.90\%&99.91\% &94.90\% &99.91\% \\
                         \bottomrule
\end{tabular}
\vspace{-0.5\intextsep}
\caption{The performances of different attacks in terms of ASR and CACC in percentage.}
\vspace{-0.5\intextsep}
\label{t:attack_results}
\end{table*}

\section{More Details Detection Results}
\label{a:detection}
We present detailed detection results of MS and baseline methods on the YELP and AG's News datasets in Tables~\ref{t:detection_yelp} and~\ref{t:detection_ag}, reporting AUC, FAR, and FRR across various attacks and models for a comprehensive robustness evaluation.
MS consistently achieves the highest AUC and maintains low FAR and FRR across most settings, demonstrating strong and stable detection performance.
Compared to STRIP and DAN, MS exhibits superior robustness against Stylebkd and Synbkd attacks, where other methods often suffer from elevated FAR. While NAS shows competitive results in some instances, its performance is less stable, particularly on complex architectures.
A key advantage of MS is its cross-architecture generalizability. It performs reliably across a wide range of models, from PLMs (such as BERT and BART) to LLMs (including LLaMA 3B and Qwen 3B). Importantly, MS does not require architecture-specific tuning, making it highly suitable for practical deployment.

\begin{table*}[ht]
\centering
\setlength{\tabcolsep}{2.0mm}
\begin{tabular}{ccccccccccccccccc}
\toprule
\multirow{2}{*}{\textbf{Attack}} & \multirow{2}{*}{\textbf{Defense}} & \multicolumn{3}{c}{\textbf{BERT}} & \multicolumn{3}{c}{\textbf{BART}} & \multicolumn{3}{c}{\textbf{LLaMA 3B}} & \multicolumn{3}{c}{\textbf{Qwen 3B}} \\ \cmidrule(lr){3-5} \cmidrule(lr){6-8} \cmidrule(lr){9-11}  \cmidrule(lr){12-14}
&&AUC& FAR& FRR&AUC&FAR&FRR&AUC&FAR& FRR&AUC& FAR& FRR \\
\midrule
\multirow{4}{*}{BadNets}&STRIP&53.73&83.28&11.83&53.14&84.34&10.16&53.20&83.21&13.96&54.70&82.95&13.36 \\
&DAN&93.51&37.64&7.03&77.29&61.89&7.40&66.69&85.08&7.20&75.40&79.08&6.96 \\
&NAS&\textbf{98.49}&\textbf{0.00}&\textbf{5.23}&96.07&30.58&4.73&88.21&83.34&\textbf{5.93}&97.43&2.27&\textbf{6.10} \\
&MS&98.10&\textbf{0.00}&6.30&\textbf{97.54}&\textbf{9.06}&\textbf{5.03}&\textbf{95.62}&\textbf{18.12}&6.23&\textbf{99.62}&\textbf{0.07}&6.53 \\ \cline{2-14}
\multirow{4}{*}{AddSent}&STRIP&52.41&85.54&10.83&53.58&84.81&11.00&51.42&94.54&\textbf{5.26}&49.36&93.14&\textbf{5.43} \\
&DAN&77.62&86.01&7.00&90.59&35.98&6.93&65.33&85.81&6.83&69.68&89.07&6.70 \\
&NAS&99.72&\textbf{0.00}&\textbf{5.16}&96.22&\textbf{0.53}&5.96&99.17&0.20&5.93&99.56&\textbf{0.00}&5.90 \\
&MS&\textbf{99.92}&\textbf{0.00}&5.53&\textbf{99.38}&2.27&\textbf{5.13}&\textbf{99.84}&\textbf{0.00}&6.86&\textbf{99.87}&\textbf{0.00}&6.53 \\ \cline{2-14}
\multirow{4}{*}{Stylebkd}&STRIP&52.10&86.34&10.86&54.82&78.48&15.26&55.68&87.21&9.70&55.17&84.41&12.06 \\
&DAN&96.61&13.06&7.60&98.55&4.20&7.70&89.24& 35.58&7.33&90.77&32.91&6.36 \\
&NAS&97.27&8.39&\textbf{5.50}&\textbf{99.98}&\textbf{0.00}&\textbf{5.13}&99.78&\textbf{0.00}&\textbf{5.96}&99.75&\textbf{0.00}&\textbf{5.80} \\
&MS&\textbf{98.38}&\textbf{5.13}&6.56&99.97&0.07&5.30&\textbf{99.90}&\textbf{0.00}&6.36&\textbf{99.86}&\textbf{0.00}&7.23 \\ \cline{2-14}
\multirow{4}{*}{Synbkd}&STRIP&50.23&93.94&\textbf{5.53}&50.95&86.74&9.33&55.60&82.55&13.06&55.16&85.81&11.16 \\
&DAN&92.72&40.77&7.03&99.65&0.20&7.33&99.11&2.00&7.70&94.23&26.38&7.16 \\
&NAS&95.84&19.19&5.86&99.70&\textbf{0.00}&\textbf{5.80}&99.99&\textbf{0.00}&\textbf{5.33}&99.78&\textbf{0.00}&\textbf{6.26} \\
&MS&\textbf{97.73}&\textbf{6.86}&6.83&\textbf{99.99}&0.07&5.86&\textbf{100}&\textbf{0.00}&6.96&\textbf{99.86}&\textbf{0.00}&8.06 \\ 
\bottomrule
\end{tabular}
\vspace{-0.5\intextsep}
\caption{Backdoor detection performance (in percentage) of MS and baselines on YELP. Best results are \textbf{bolded}.}
\vspace{-0.5\intextsep}
\label{t:detection_yelp}
\end{table*}

\begin{table*}[ht]
\centering
\setlength{\tabcolsep}{2.0mm}
\begin{tabular}{ccccccccccccccccc}
\toprule
\multirow{2}{*}{\textbf{Attack}} & \multirow{2}{*}{\textbf{Defense}} & \multicolumn{3}{c}{\textbf{BERT}} & \multicolumn{3}{c}{\textbf{BART}} & \multicolumn{3}{c}{\textbf{LLaMA 3B}} & \multicolumn{3}{c}{\textbf{Qwen 3B}} \\ \cmidrule(lr){3-5} \cmidrule(lr){6-8} \cmidrule(lr){9-11}  \cmidrule(lr){12-14}
&&AUC& FAR& FRR&AUC&FAR&FRR&AUC&FAR& FRR&AUC& FAR& FRR \\
\midrule
\multirow{4}{*}{BadNets}&STRIP&53.14&82.42&14.12&52.44&82.89&14.05&52.21&85.35&12.07&53.81&78.49&18.07 \\
&DAN&86.86&70.93&4.87&99.76&\textbf{0.00}&5.33&87.46&32.90&\textbf{4.75}&86.37&36.98&5.33 \\
&NAS&95.19&\textbf{29.65}&4.91&80.08&83.25&\textbf{4.75}&92.12&77.14&4.82&95.85&32.90&4.62 \\
&MS&\textbf{93.74}&39.54&\textbf{4.25}&\textbf{99.97}&\textbf{0.00}&5.01&\textbf{99.90}&\textbf{0.00}&5.00&\textbf{98.91}&\textbf{0.26}&\textbf{4.46} \\ \cline{2-14}
\multirow{4}{*}{AddSent}&STRIP&50.78&88.09&10.00&53.11&80.19&15.41&51.75&88.25&10.12&52.73&84.68&13.11 \\
&DAN&81.24&82.72&4.76&89.54&41.02&\textbf{4.97}&86.76&34.67&\textbf{4.61}&86.87&42.09&4.95 \\
&NAS&\textbf{96.96}&\textbf{17.33}&5.33&81.68&83.19&5.08&95.34&60.74&4.78&90.61&95.35&5.15 \\
&MS&96.63&18.61&\textbf{4.62}&\textbf{99.11}&\textbf{3.68}&5.39&\textbf{99.95}&\textbf{0.00}&5.04 &\textbf{99.94}&\textbf{0.00}&\textbf{3.80 }\\ \cline{2-14}
\multirow{4}{*}{Stylebkd}&STRIP&50.36&93.11&6.51&51.41&84.91&14.34&53.84&80.84&15.96&54.46&79.10&16.42 \\
&DAN&95.41&21.19&\textbf{4.49} &98.33&6.16&4.67&86.69&\textbf{0.00}&100 &90.23&\textbf{0.00}&100 \\
&NAS&\textbf{98.22}&\textbf{5.61}&4.75&95.14&18.86&4.92&89.61&36.37&\textbf{4.78}&93.30&32.19&5.04 \\
&MS&97.36&6.04&4.71&\textbf{97.66}&\textbf{5.98}&\textbf{4.53}&\textbf{95.14}&18.79&\textbf{4.78}&\textbf{97.10}&8.79&\textbf{4.11 }\\ \cline{2-14}
\multirow{4}{*}{Synbkd}&STRIP &52.46&79.56&16.84&54.28&72.63&21.65&55.87&67.51&24.50 &55.85&70.23&22.25 \\
&DAN&99.89&\textbf{0.00}&100&99.94&\textbf{0.00}&100&99.93&\textbf{0.00}&100&99.91&\textbf{0.00}&100 \\
&NAS&99.35&0.79&5.64&99.90&0.11&\textbf{4.82}&98.62&0.09&5.16&96.71&13.16&4.78 \\
&MS&\textbf{99.62}&0.30&\textbf{4.45}&\textbf{99.96}&0.09&5.11&\textbf{99.97}&0.09&\textbf{4.63}&\textbf{99.95}&0.09&\textbf{4.16} \\
\bottomrule
\end{tabular}
\caption{Backdoor detection performance (in percentage) of MS and baselines on AG's News. Best results are \textbf{bolded}.}
\label{t:detection_ag}
\end{table*}

\section{Ablation Study on the Number of Selected Layers}
\label{a:layers}
This section investigates the influence of the hyperparameter $k$, the number of layers selected using the top-$k$ strategy, on the detection performance.
We adopt $k = 3$ for both PLMs and LLMs in our experiments. To assess the impact of $k$, we conduct experiments at discrete values: $k = 1, 3, 5$, and $7$.

\begin{figure}
    \centering
    \includegraphics[width=\linewidth]{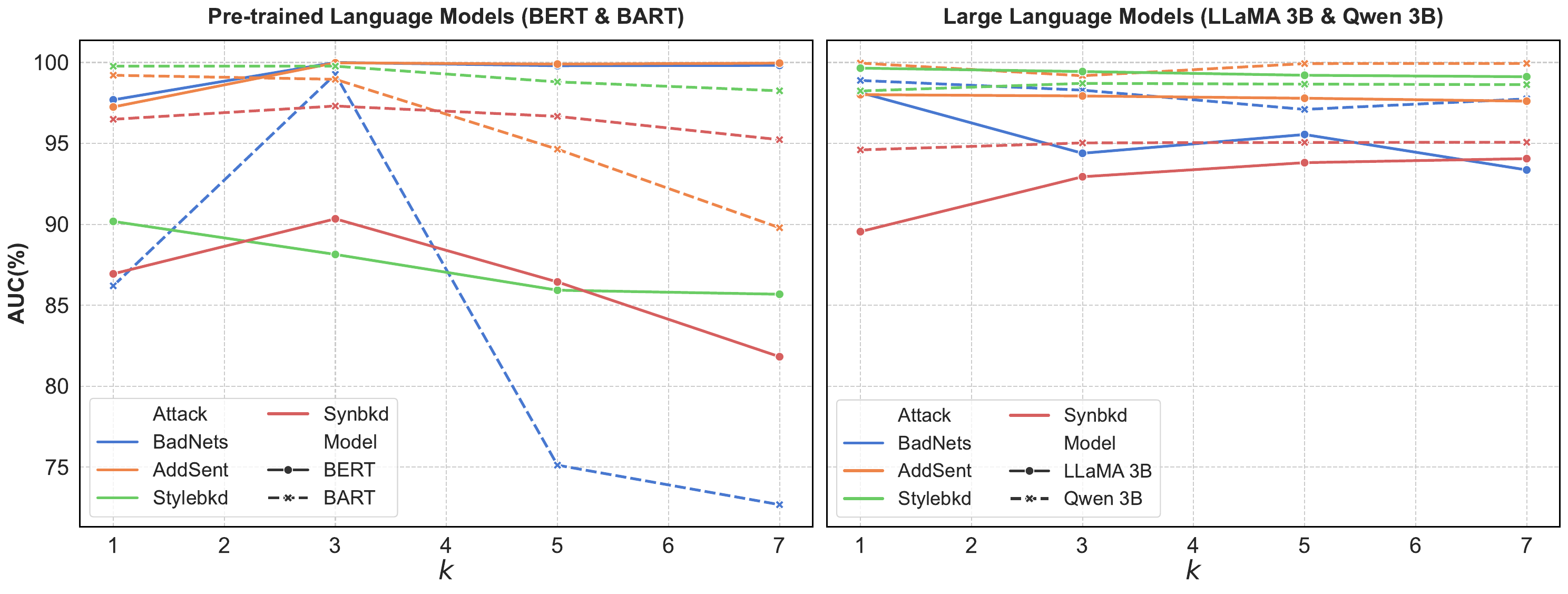}
    \caption{Impact of varying $k$ on the detection performance of PLMs (\textbf{left}) and LLMs (\textbf{right}) on the SST-2 dataset.}
    \label{f:topk}
\end{figure}

As illustrated in Figure~\ref{f:topk}, increasing $k$ generally improves AUC scores up to a certain point, after which the performance slightly declines. Specifically, for PLMs such as BERT and BART, the AUC tends to peak or stabilize around $k = 3$, with BERT showing robust results at this setting. A similar trend is observed in LLaMA 3B and Qwen 3B, where $k = 3$ offers competitive performance across different attack scenarios. These results suggest that selecting the top 3 most informative layers balances effectiveness and efficiency. Therefore, we set $k = 3$ as the default configuration in all main experiments to ensure high detection performance and computational efficiency.

\section{Ablation Study on the Fusion Weigh}
\label{a:alpha}
In computing the final anomaly score $S_{\text{final}}(x)$, we introduce a fusion hyperparameter $\alpha$ to balance the static and dynamic anomaly score. To investigate the impact of $\alpha$, we conduct an ablation study by varying its value from 0.5 to 1.0 in increments of 0.1.

\begin{figure}
    \centering
    \includegraphics[width=\linewidth]{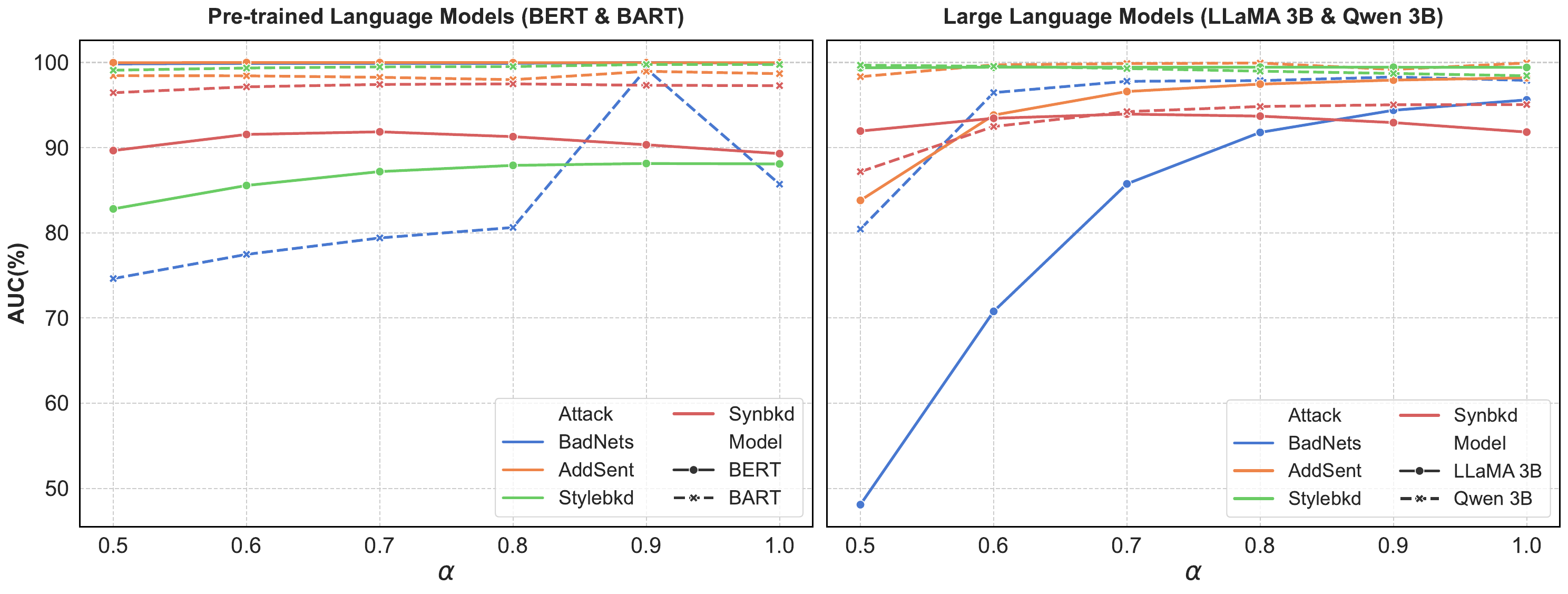}
    \caption{Impact of varying $\alpha$ on the detection performance of PLMs (\textbf{left}) and LLMs (\textbf{right}) on the SST-2 dataset.}
    \label{f:alpha}
\end{figure}

As illustrated in Figure~\ref{f:alpha}, the detection performance is sensitive to the choice of $\alpha$. When $\alpha = 0.9$, most models achieve best or close-to-best AUC scores, especially BART under the BadNets attack and Qwen 3B across multiple settings. In contrast, lower values of $\alpha$ (e.g., 0.5 or 0.6) often result in degraded performance, likely due to noise in the dynamic signal. Overemphasizing $S_{\text{SS}}$ in the final score may amplify this noise and hurt detection accuracy. This confirms that $\alpha = 0.9$ provides a good balance between static and dynamic signals, and is thus adopted in our main experiments.

\section{Ablation Study on the Number of Clean Samples}
\label{a:cleandata}
\begin{figure}[ht]
    \centering
    \includegraphics[width=\linewidth]{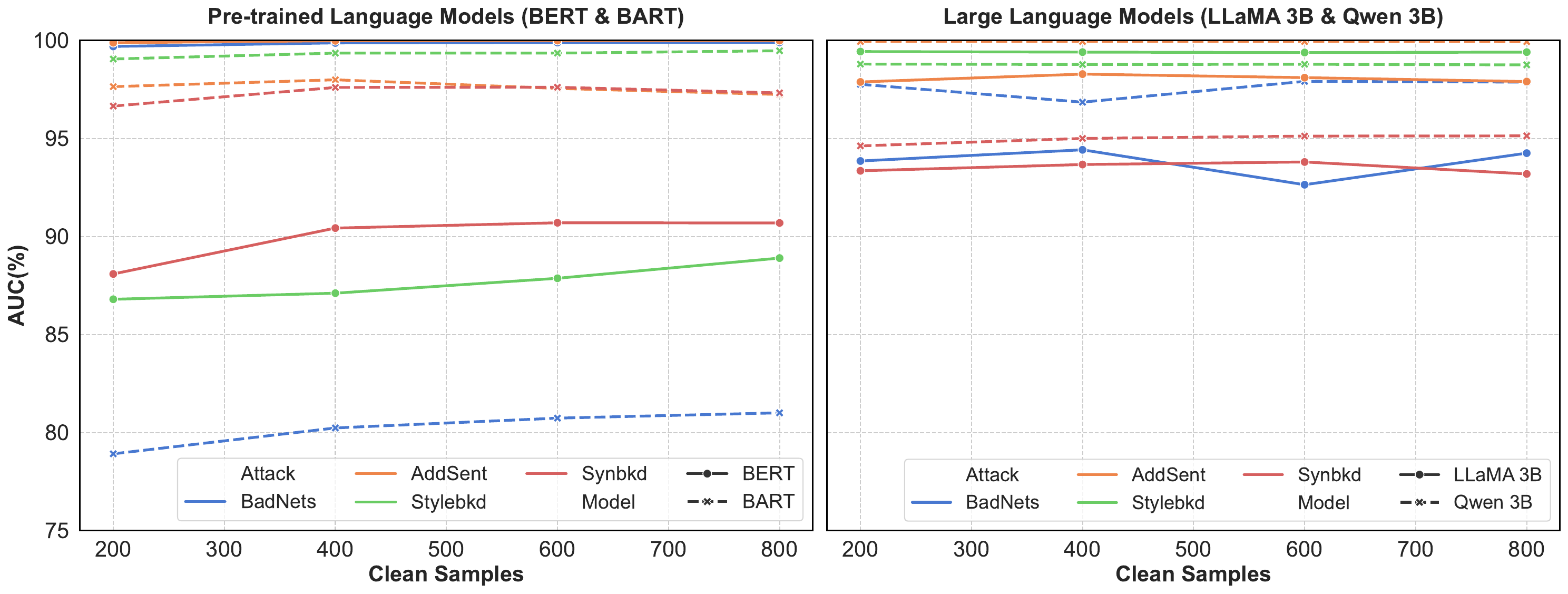}
    \caption{Impact of clean samples on the detection performance of PLMs (\textbf{left}) and LLMs (\textbf{right}) on the SST-2 dataset.}
    \label{f:clean_sample}
\end{figure}

To evaluate the robustness of our method with respect to the amount of clean data available, we conduct an ablation study by varying the number of clean samples.
Specifically, we evenly split the limited clean dataset $\mathcal{D}$ into two subsets: one for calibration $\mathcal{D}_{\text{calib}}$ and the other for validation $\mathcal{D}_{\text{valid}}$.

As illustrated in Figure~\ref{f:clean_sample}, our MS maintains consistently high AUC scores across a wide range of sample sizes for both PLMs and LLMs, demonstrating strong robustness to the quantity of clean data. Even with as few as 200 clean samples, the detection performance remains competitive, highlighting our MS's efficiency and practicality in low-resource scenarios.
Notably, the BART model under the BadNets attack shows slightly lower AUC than LLMs. This may be attributed to LLMs' superior representation and generalization capabilities, which allow them to construct more stable feature distributions from limited data.

\section{Details of Adaptive Attacks}
\label{a:adaptive}

To implement the adaptive attack, we employ a strategy based on feature-level regularization~\cite{Zhao0XDWL22,ZhongQZ22}. The core objective is to force the latent representations of poisoned samples to mimic those of clean samples, making them indistinguishable in the model's feature space.

We introduce a regularization loss term, $\mathcal{L}_{\text{ce}}$, which minimizes the distance between the feature representations of poisoned and clean samples across all $L$ layers of the model. This loss is defined as:
\begin{equation}
\mathcal{L}_{\text{reg}} = \sum_{1 \le i \le L} \lVert f_i^{\text{poisoned}} - f_i^{\text{clean}} \rVert_2,
\end{equation}
where $f_i^{\text{poisoned}}$ and $f_i^{\text{clean}}$ are the feature representation vectors of the poisoned and clean samples at layer $i$, respectively, and $\lVert \cdot \rVert_2$ denotes the Euclidean distance (L2-norm). The clean samples are specifically chosen to have the same ground-truth label as the backdoor's target label.

The final training objective combines this regularization term with the standard cross-entropy loss $\mathcal{L}_{ce}$:
\begin{equation}
\mathcal{L} = \mathcal{L}_{\text{ce}} + \alpha \mathcal{L}_{\text{reg}},
\end{equation}
where $\alpha$ is a hyperparameter that balances the primary task of backdoor injection with the secondary goal of feature concealment. Consistent with prior work~\cite{dan,badacts}, we set $\alpha$=250 in our experiments to ensure the regularization is sufficiently strong.

\section{Analysis of DUP’s Efficiency Advantages}
\label{a:efficiency}
This section analyzes the computational and memory efficiency of the proposed DUP framework, highlighting three core design choices that contribute to its practical applicability.

\noindent \paragraph{Top-$k$ Layer Selection for Efficient Detection.}
Unlike many existing feature-space defenses, which rely on representations from all layers, DUP adopts a top-$k$ selection strategy focusing only on the most discriminative layers (e.g., $k=3$ in our experiments). This significantly reduces the computational overhead by limiting costly operations such as Mahalanobis distance calculation to a small subset of informative layers rather than the full model. This design streamlines detection by avoiding redundant analysis of less valuable layers.

\noindent \paragraph{Parameter-Efficient Unlearning via LoRA.}
Conventional purification methods typically require full-parameter fine-tuning, which is computationally demanding, particularly for large-scale models. DUP leverages Low-Rank Adaptation (LoRA) to perform unlearning in a parameter-efficient manner. By freezing the base model and updating only lightweight LoRA adapters, DUP substantially reduces the number of trainable parameters, resulting in faster optimization and lower memory usage during training.

\noindent \paragraph{Memory-Efficient Self-Contained Distillation.}
Some prior methods require loading an external clean model to guide the purification process, leading to high memory consumption. In contrast, DUP employs a self-contained distillation scheme where the backdoored model serves as its teacher, and a LoRA-injected copy acts as the student. This eliminates the need for additional clean models, thereby reducing peak GPU memory usage, which is particularly beneficial when defending resource-intensive LLMs.

These efficiency-oriented design choices make DUP a practical and scalable defense framework, particularly suited for real-world deployment under limited computational and memory budgets.

\end{document}